\documentclass[a4paper,fleqn,usenatbib]{mnras}

\usepackage[T1]{fontenc}
\usepackage{ae,aecompl}

\usepackage{graphicx}	
\usepackage{amsmath}	
\usepackage{amssymb}	

\title[Accretion of Super-Earth Envelopes]{The Imprint of the Protoplanetary Disk in the Accretion of Super-Earth Envelopes}

\author[Ali-Dib, Cumming, \& Lin]{
Mohamad Ali-Dib$^{1}$\thanks{E-mail: malidib@astro.umontreal.ca}, Andrew Cumming$^{2}$, and Douglas N. C. Lin$^{3,4}$ \\
$^{1}$ Institut de recherche sur les exoplan\`etes, D\'epartement de physique, Universit\'e de Montr\'eal,\\
2900 boul. \'Edouard-Montpetit, Montr\'eal, Quebec, H3T 1J4, Canada\\
$^{2}$Department of Physics and McGill Space Institute, McGill University, 3600 rue University, Montreal, QC, H3A 2T8, Canada\\
$^{3}$Department of Astronomy and Astrophysics, University of California, Santa Cruz, CA 95064, USA\\
$^{4}$Institue for Advanced Studies, Tsinghua University, Beijing, China
}

\date{Accepted XXX. Received YYY; in original form ZZZ}

\pubyear{2019}

\begin{document}
\label{firstpage}
\pagerange{\pageref{firstpage}--\pageref{lastpage}}
\maketitle

\begin{abstract}
Super-Earths are by far the most dominant type of exoplanet, yet their formation is still not well understood. In particular, planet formation models predict that many of them should have accreted enough gas to become gas giants. Here we examine the role of the protoplanetary disk in the cooling and contraction of the protoplanetary envelope. In particular, we investigate the effects of {1) the thermal state of the disk as set by the relative size of heating by accretion or irradiation, and whether its energy is transported by radiation or convection,} and 2) advection of entropy into the outer envelope by disk flows that penetrate the Hill sphere, as found in 3D global simulations. We find that, at 5 and 1 AU, this flow at the level reported in the non-isothermal simulations where it penetrates only to $\sim$ 0.3 times the Hill radius has little effect on the cooling rate since most of the envelope mass is concentrated close to the core, and far from the flow. On the other hand, at 0.1 AU, the envelope quickly becomes fully-radiative, nearly isothermal, and thus cannot cool down, stalling gas accretion. This effect is significantly more pronounced in convective disks, leading to envelope mass orders of magnitude lower.
Entropy advection at 0.1 AU in either radiative or convective disks could therefore explain why super-Earths failed to undergo runaway accretion. These results highlight the importance of the conditions and energy transport in the protoplanetary disk for the accretion of planetary envelopes.

\end{abstract}




\section{Introduction}
One of the main discoveries of exoplanet finder \textit{Kepler} is the presence of a very large number of planets in the 1-4 R$_\oplus$ radii range, referred to as super-Earths (SEs), or sub-Neptunes \citep{fresin}. While some of these objects are consistent with an Earth-like composition, others are of lower mean density, implying that extended H/He envelopes are needed to explain their mass and radius \citep{lopez,weiss}. Neither super-Earths or sub-Neptunes have analogues in the solar system, and both probably represent an intermediate stage of planet formation, having not accreted enough gas to transition into giant planets \citep{bitsch}. They are hence a laboratory to test planet formation models that have yet to converge on an explanation of their origins. 

A central question is why do super-Earths significantly outnumber Neptunian and Jovian mass planets, or why didn't they accrete more gas? In the context of core accretion models \citep{pollack}, the key physical process to examine is the phase in which the gas envelope is joined smoothly to the disk, and gas accretes slowly and hydrostatically onto the core. The rate of gas accretion is regulated by the atmosphere's cooling efficiency, as loss of thermal energy causes the envelope to contract, driving accretion.
Many works have examined this process in the context of super-Earth formation. \cite{lee1} and \cite{lee2} for example showed that a 5-10 M$_\oplus$ core at 0.1 AU can reach gas runaway accretion before the dissipation of the disk in a large swath of parameter space. 
{They hence proposed that super-Earths either form in a low gas to dust ratio environment, in which cooling of the envelope is slow due to high opacities {(although, as found by \cite{lee2016}, high enough dust abundance will increase the mean molecular weight, decreasing cooling time),} or through collisional mergers of protoplanets where the final mass doubling (the final merger that creates 5-10 M$_\oplus$ cores) occurs sufficiently late in disk evolution so that there is not enough time nor enough gas for super-Earth cores to undergo runaway accretion. {Alternatively, \cite{kite} proposed that, for planets around this mass, the pressure at the base-of-atmosphere is high enough to sequester the atmosphere into magma core, quenching further growth of gaseous envelope.}}

Another possible solution to this problem was put forward initially by \cite{ormel1} and \cite{ormel2}, who conducted hydrodynamic simulations of a core embedded in a disk. They found that, for isothermal conditions, the envelope-disk system behaves as an open system, with continual exchange of gas between the Hill sphere and the disk. Flow of disk gas into the planetary envelope continuously replenishes it with higher entropy gas, slowing down atmospheric cooling. Here, we refer to this process as ``entropy advection''. Similar results were found in adiabatic simulations \citep{popovas}. Models including radiative cooling \citep{lamb1, kuro}, however, found the flow to penetrate only to $\sim$ 0.3 R$_{\text{Hill}}$, with the gas inside this radius being hydrostatically bound to the core, significantly muting the effect. These global simulations however were conducted only for few orbital periods due to their numerical cost. 

In this paper, we investigate the influence of entropy advection on the long term evolution of planetary envelopes during the slow gas accretion phase. We adopt 1D cooling models of the planetary envelope, that can follow its evolution over long timescales, but modify the outer boundary condition to include the effect of advective winds. 
The main goal is to quantify how much entropy advection at the levels reported by the non-isothermal hydrodynamic simulations affect this phase. We moreover investigate the range of entropy expected for the disk gas. In particular, convective transport of energy in the disk leads to a larger entropy than in a radiative disk. Both of these aspects change the entropy at the outer boundary of the envelope, and therefore the envelope structure and cooling rate. While our model is simplified, especially in contrast with 3D global hydrodynamic simulations {that consistently account for important effects such as rotation \citep{fung}}, it allows us to follow envelope cooling all the way to runaway accretion and investigate further the role of this mechanism in the formation of super-Earths.

We start in \S 2 with a discussion of the expected conditions in the disk near a super-Earth progenitor, and compare with conditions assumed in previous work. In \S 3, we use 1D models of the cooling envelope to follow the growth of the planet under different disk conditions, including a modified outer boundary condition that approximates the effect of entropy advection. We conclude and discuss the implications for super-Earth formation in \S 4.

\begin{figure*}
\begin{centering}
	\includegraphics[scale=0.21]{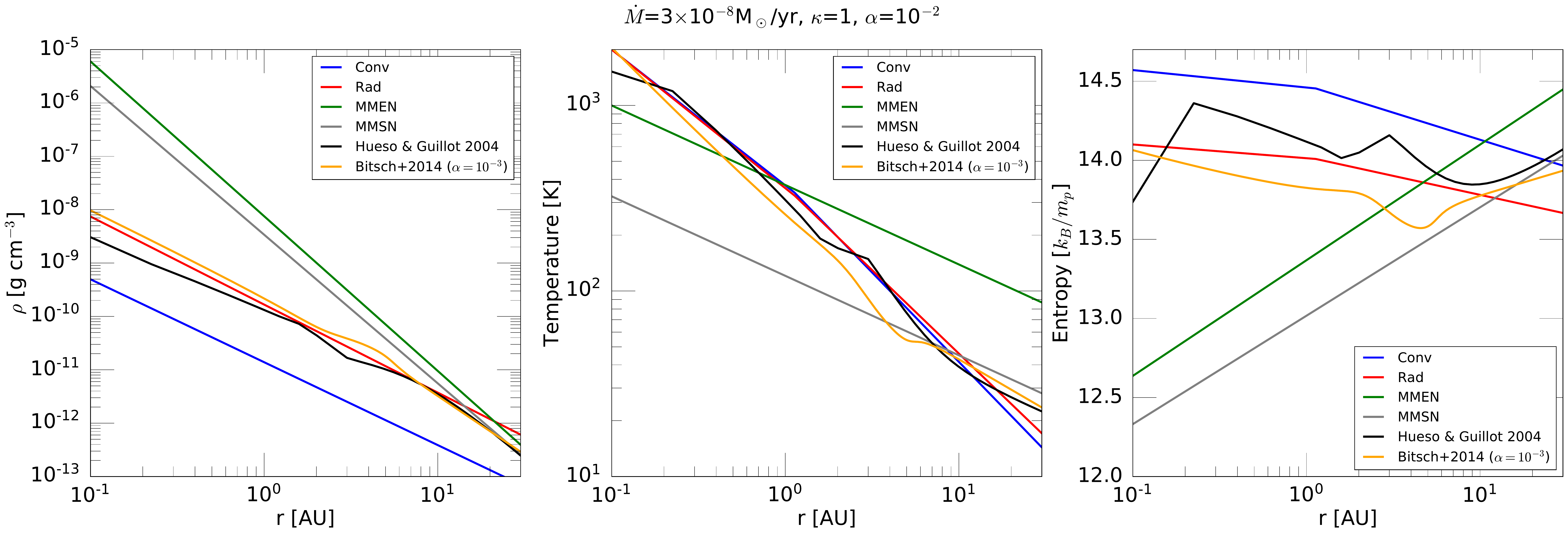}
   \caption{The density (left), temperature (center), and entropy (right) profiles of our radiative (red) and convective (blue) alpha-disk models for an accretion rate $\dot M=3\times 10^{-8}\ M_\odot\ {\rm yr^{-1}}$. For comparison, we show other disk models from the literature: accreting disk models from  \citep{hueso} and \citep{bitschdisk}, and the minimum mass solar nebula (MMSN) \citep{piso} and minimum mass extrasolar nebula (MMEN) \citep{mmen}. All the accreting disk models have similar temperature and density profiles, with the one difference that the convective disk has a lower density and higher entropy than the radiative disks. The passive disks have much higher densities and lower temperatures and entropy in the inner parts of the disk within 1 AU. }
    \label{fig:disk1}
    \end{centering}
\end{figure*}

\section{Conditions in the Protoplanetary Disk}

In the initial growth of a planetary envelope, when the mass of the envelope is smaller than the mass of the solid core, the envelope extends out to a distance comparable to the Bondi or Hill radius, where it smoothly joins onto the disk. The accretion rate of gas onto the core, or equivalently the growth rate of the envelope, is set by how quickly the envelope can cool. {This is ``Phase 2'' of the core-accretion model of planet formation \citep{pollack,morda}}.  In this phase, models use the disk density and temperature as an outer boundary condition when calculating the cooling rate of the envelope. 

A common choice is to take the temperature profile of the disk as set by irradiation from the central star \citep{chiang} combined with an assumed disk density profile. For example, \cite{lee1} take the minimum-mass extrasolar nebula (MMEN) from \cite{mmen}, with density and temperature
\begin{eqnarray}
    \rho_{\rm MMEN} &=& 7.6\times 10^{-9}\ {\rm g\ cm^{-3}}\ r_{\rm AU}^{-2.9}\\
    T_{\rm MMEN} &=& 373\ {\rm K}\ r_{\rm AU}^{-3/7},
\end{eqnarray}
where $r_{\rm AU}$ is the distance from the star measured in AU. \cite{piso} use a minimum-mass solar nebula (MMSN) with similar power law slopes for the temperature and density profiles, but colder ($\approx 120\ {\rm K}$ at 1 AU), and less dense by a factor of $2$--$3$ \citep{mmen}. Another example is \cite{rafikov2006} who used a similar profile.

The conditions in the disk could, however, be quite different if the disk is actively accreting, particularly close to the star where viscous heating dominates stellar irradiation \citep{garaud2007}. To model this case, we use a standard steady-state thin disk \citep{pringle}. For a given accretion rate $\dot M$, the surface temperature $T_s$ and column density $\Sigma$ in the disk are given by 
\begin{equation}
    \sigma T_s^4 = {3GM_\star \dot M\over 8\pi r^3}; \hspace{1cm}
    \Sigma = {\dot M\over 3\pi \nu} = {\dot M\over 3\pi \alpha \Omega H^2},
\end{equation}
where $r$ is the distance from the star, $\dot{M}$ the accretion rate of the disk onto the star, $M_\star$ is the stellar mass, $\Omega = (GM_\star/r^3)^{1/2}$, and we use the alpha prescription for viscosity in the form $\nu = \alpha \Omega H^2$. Unless otherwise specified we adopt values $\alpha=10^{-2}$, $\dot M = 3\times 10^{-8} M_\odot\ {\rm yr^{-1}}$, and $M_\star = 1 M_\odot$ for the models in this paper. Note that the disk conditions depend on $M_\star$ and $r$ only through $\Omega$, so that our results can be rescaled to different stellar masses by changing $r$ to keep the same orbital period. {In this work we assume the disk's accretion rate to be constant. In reality however this accretion rate will decrease with time, cooling down the envelope's outer boundary. }

For the outer boundary of our envelope models, we need the temperature $T_d$ and density $\rho_d$ near the disk midplane. The density is given by 
\begin{equation}
    \rho_d = {\Sigma\over 2H},
\end{equation}
where the disk scale height from hydrostatic balance in the vertical direction is
\begin{equation}
    H = \beta {c_s\over \Omega}
\end{equation}
with $c_s$ the sound speed in the disk given by $c_s^2 = \gamma P/\rho$, and $\beta$ a parameter of order unity that depends on the vertical profile of temperature and density in the disk. Assuming an ideal gas equation of state $P=\rho k_B T/\mu$, where $\mu$ is the mean molecular weight,
we find that the central density and temperature are related by
\begin{equation}\label{eq:rhoc_Tc_relation}
    \rho_d = {\dot M \Omega^2\over 6\pi \alpha \beta^3}\left({\mu\over \gamma k_B}\right)^{3/2}{1\over T_d^{3/2}}.
\end{equation}
We assume constant $\gamma=1.4$ and $\mu=2.34\,m_p$.

The disk temperature $T_d$ is determined by the vertical energy transport in the disk. Here, we consider two models. The first is a radiative disk with constant opacity $\kappa$. In that case, the radiative diffusion equation gives
\begin{equation}
T_{d,\mathrm{rad}}^4 = \tau T_s^4 = \left({\kappa\Sigma\over 2}\right)T_s^4    
\end{equation}
where $\tau$ is the optical depth through the disk. In this case, we set $\beta=1$, from which it follows that 
\begin{eqnarray}
    T_{d,\mathrm{rad}} &=& T_s^{4/5} \left({\kappa\mu\Omega \dot M\over 6\pi \alpha k_B \gamma}\right)^{1/5}\nonumber\\
    &=& 373\ {\rm K}\  r_{\rm AU}^{-9/10} \alpha_{-2}^{-1/5} \dot M_{-7.5}^{2/5}\nonumber\\ &&\times\left({M_\star\over M_\odot}\right)^{3/10}\left({\kappa\over {\rm cm^2\ g^{-1}}}\right)^{1/5}
\end{eqnarray}
where $\alpha_{-2}=\alpha/0.01$ and $\dot M_{-7.5} = \dot M/10^{-7.5}\ M_\odot\ {\rm yr^{-1}}$. The density is then 
\begin{eqnarray}
    \rho_{d, \mathrm{rad}} &=& 1.7\times 10^{-10}\ {\rm g\ cm^{-3}}\ r_{\rm AU}^{-33/20}\alpha_{-2}^{-7/10}\dot M_{-7.5}^{2/5} \nonumber\\
    &&\times \left({M_\star\over M_\odot}\right)^{11/20}\left({\kappa\over {\rm cm^2\ g^{-1}}}\right)^{-3/10}.
\end{eqnarray}
The second case is a convective disk \citep{lin1} in which the convection is efficient so that the thermal profile corresponds to an adiabat. Then
\begin{equation}
    T_{d,\mathrm{conv}} = T_s \left({\rho_c\over\rho_s}\right)^{\gamma-1},
\end{equation}
where the density at the surface of the disk is $\rho_s \approx 1/\kappa H$. Since $P\propto \rho^\gamma$ along the adiabat, hydrostatic balance in the vertical direction gives $\beta^2 = 2/(\gamma-1)$. Using equation (\ref{eq:rhoc_Tc_relation}) for $\rho_c$, we find 
\begin{eqnarray}
    T_{d,\mathrm{conv}} &=& T_s^{1/\gamma} \left({\kappa\mu\Omega \dot M\over 6\pi \alpha k_B \gamma \beta^2}\right)^{(\gamma-1)/\gamma}\nonumber\\
&=& 391\ {\rm K}\ r_{\rm AU}^{-27/28} \dot M_{-7.5}^{13/28}\alpha_{-2}^{-2/7}\nonumber\\&&\times \left({M_\star\over M_\odot}\right)^{9/28} \left({\kappa\over {\rm cm^2\ g^{-1}}}\right)^{2/7}
\end{eqnarray}
where we take $\gamma=1.4$. The density is
\begin{eqnarray}
    \rho_{d, \mathrm{conv}} &=& 1.4\times 10^{-11}\ {\rm g\ cm^{-3}}\ r_{\rm AU}^{-87/56}\alpha_{-2}^{-4/7}\dot M_{-7.5}^{17/56}\nonumber\\
    &&\times \left({M_\star\over M_\odot}\right)^{29/56}\left({\kappa\over {\rm cm^2\ g^{-1}}}\right)^{-3/7}.
\end{eqnarray}
In both cases, we approximate the effect of dust sublimation, which significantly lowers the opacity and therefore the temperature gradient in the disk \citep{DAlessio}, by limiting the temperature to be $T_d<2000\ {\rm K}$.

We show the temperature and density profiles for the radiative and convective disks in Figure \ref{fig:disk1}, compared with two more detailed models from the literature \citep{hueso,bitschdisk}.
To generate the profiles from the code provided by \cite{hueso}, we start with a 0.01 M$_\odot$ star surrounded by a 1 M$_\odot$ cloud at 10 K. We evolve the system for $\sim$ 0.5 Myr until the mass and accretion rate of the disk becomes comparable to ours. To generate the disk profiles from the code provided by \cite{bitschdisk}, we use the default parameters of the fit. We see that the density and temperature profiles have mostly similar slopes. The effect of dust sublimation can be seen in the flattening of the disk temperature profile for $r\lesssim 0.2\ {\rm AU}$. We also compare our accreting disk models with the passive disk models MMEN \citep{mmen} and MMSN \citep{piso}. The accreting disks are significantly hotter and less dense than the passive disks in the inner parts of the disk $r\lesssim 1\ {\rm AU}$.

We also show the entropy of the disk in Figure \ref{fig:disk1}. Entropy will be important for our models of the disk flow that penetrate the Hill sphere, since the flow timescale is short enough that the flow is adiabatic. The entropy at the outer boundary is also a relevant quantity for comparison with the interior of the planetary envelope, which is convective in many cases and therefore also follows an adiabat. We find that the entropy in the inner part of the disk is significantly larger in accreting models than the passive disk models. The trend of entropy with radius is also opposite in the two cases, with entropy decreasing outwards for accreting disks, but increasing outwards for passive disks. This is because the MMEN has a shallower temperature profile ($T_d\propto r^{-3/7}$) and steeper density profile (approximately $\rho_d\propto r^{-3}$), giving approximately $P/\rho^\gamma\propto r^{4/5}$, compared to our radiative or convective models, which both have scalings close to $T_d\propto r^{-1}$ and $\rho_d\propto r^{-3/2}$, giving $P/\rho^\gamma\propto r^{-2/5}$.

\section{Cooling models including entropy advection}

In this section, we model the evolution of the protoplanetary atmosphere as it cools and accretes, and explore the effects of entropy advection and different disk models. Details of the cooling models are given in \S 3.1, including a modification of the outer boundary condition that approximately takes into account entropy advection, and then we present our results in \S 3.2.


\subsection{Atmospheric cooling model}

\subsubsection{Structure equations and evolution}

To model the slow-phase gas accretion, we first define the standard atmospheric structure equations as:

\begin{equation}
\frac{dm}{dr} = 4\pi r^2 \rho(r)
\end{equation} 

\begin{equation}
\frac{dP}{dr} = -\frac{Gm}{r^2}\rho(r)
\end{equation}

\begin{equation}
\frac{dT}{dr} = \nabla \frac{T}{P} \frac{dP}{dr}
\end{equation}
where $\nabla = \text{min}(\nabla_{\mathrm{ad}}, \nabla_{\mathrm{rad}})$ with $\nabla_{\mathrm{ad}}$ the adiabatic gradient:

\begin{equation}
\nabla_{\mathrm{ad}} \equiv\left(\frac{d \ln T}{d \ln P}\right)_{\mathrm{ad}} = \frac{\gamma - 1}{\gamma}
\end{equation}
and $\nabla_{\mathrm{rad}}$ the radiative gradient:

\begin{equation}
\nabla_{\mathrm{rad}} \equiv \frac{3 \kappa P}{64 \pi G m \sigma T^{4}} L
\end{equation}
where $L$ is the envelope's luminosity, and $\kappa$ is the opacity that we define following \cite{belllin} as:
\begin{equation}
\kappa=\kappa_{i} \rho^{a} T^{b}
\end{equation}
where we take into account all of the different opacity sources used in that paper, including gas, silicates, ice, and H- scattering opacity that becomes relevant at high temperatures. We moreover artificially suppress grains opacity by a {factor 10 to account for possible grain growth and settling in the atmosphere \citep{ormel2014}}. Finally we close the system with the ideal gas equation of state $P=\rho_g k_B T/\mu$. We solve these equations by integrating inwards from the outer boundary at R$_{\rm out}$ to the core.



We replace the envelope's time dependent energy equation by the cooling treatment of \cite{piso} where we construct a large series of steady state models with increasing envelope mass by integrating the equations above, and then connect them by the calculating the timestep necessary to jump between any two adjacent steady states as: 

\begin{equation}
\Delta t = \frac{-\Delta E}{\langle L \rangle}
\end{equation}
where brackets indicate a quantity's mean value between two states, and $\Delta$ indicate their difference. $E$ is the envelope's total (gravitational and internal) energy: $E=E_{G}+U$ where
\begin{equation}
E_{G}=-\int_{M_{\mathrm{c}}}^{M} \frac{G m}{r} d m
\end{equation}
and
\begin{equation}
U=\int_{M_{\mathrm{c}}}^{M} u\,dm
\end{equation}
with
\begin{equation}
u=C_{V} T=\frac{k_B}{\mu}\left(\nabla_{\mathrm{ad}}^{-1}-1\right) T
\end{equation}
These terms are evaluated at the radiative-convective boundary level, since a fundamental assumption in \cite{piso} is that the luminosity is constant in the outer radiative zone. Note that we have ignored the two surface terms in \cite{piso} ($ \langle e \rangle \Delta M - \langle P \rangle \Delta V$), since both them and \cite{lee1} found them to be negligible except at runaway accretion where the mass accretion term increases. We checked that this applies to our model as well.

{Our model consistently includes the atmosphere's self gravity. In practice, we start calculating each snapshot by choosing a value for the envelope's mass. The eigenvalue of the problem is the luminosity $L$, where the correct value gives an integrated envelope mass equal to the value we chose initially. 
The free parameters of our atmosphere model are hence the disk temperature and density (controlled simultaneously through the semi major axis), $R_{\text{adv}}$, and the core mass $M_c$.   }

\subsubsection{Outer boundary condition and model for entropy advection}

Standard models of planet formation place the outer boundary of the planetary envelope at either the Hill sphere or Bondi radius, whichever is smaller, i.e. $R_{\rm out}$=min($R_H$,$R_B$) (e.g.~\citealt{bodenheimer86}). 
At that location, the density and temperature of the gas are set to be the same as the conditions in the disk ($\rho_d$ and $T_d$ from section 2, as summarized in Table \ref{table}). The Hill radius is $R_H=a(M_p/3M_\star)^{1/3}$, and the Bondi radius is $R_B=GM_p/c_s^2$ where $c_s^2=\gamma k_BT_d/\mu$ gives the sound speed in the disk. Their ratio is
\begin{eqnarray}
    {R_B\over R_H} &=& 3\left({R_H\over H}\right)^2\nonumber\\
    &=&0.63\ a_{\rm AU}^{-1} \left({T_d\over 400\ {\rm K}}\right)^{-1}\left({M_p\over 10\,M_\oplus}\right)^{2/3}\left({M_\star\over M_\odot}\right)^{1/3},
\end{eqnarray}
where $a$ is the distance from the star and $H=c_s/\Omega$ is the disk scale height.
Note that for our $\alpha$ disks, the quantity $aT_d$ is approximately constant, and so $R_B/R_H$ depends weakly on distance from the star.

To take into account the effects of entropy advection, we modify the outer boundary condition by first moving the outer boundary inwards to a new location $R_{\rm adv}<R_{\rm out}$. The ratio $R_{\rm adv}/R_{\rm out}$ is a parameter of the model that describes how far the disk flows penetrate into the envelope. Since the region $R_{\rm adv}<r<R_{\rm out}$ is continually replenished with disk gas on a faster timescale than the thermal timescale, we expect the gas in this region to be adiabatic, with the disk's entropy. We therefore fix the entropy of the outer boundary at $r=R_{\rm adv}$ to be the same as the disk entropy. We then integrate inwards as before, but while imposing $\nabla = \nabla_{\mathrm{ad}}$ for $R \geq R_{\text{adv}}$. We are hence assuming hydrostatic equilibrium in the radial direction in the outer advective envelope, while also accounting for the exchange of material with the disk.

We can derive an analytic expression for the increase in temperature in the entropy advection region if we neglect self-gravity of the envelope, ie. assume that the envelope mass is small compared to the core mass (note that self-gravity is included in our numerical solutions). We take the gas to be adiabatic in this region, $P=K\rho^\gamma$, and assume ideal gas as before, $P=\rho k_BT/\mu$. With these assumptions, the equation of hydrostatic balance $dP/dr = -\rho GM_p/r^2$ can be integrated inwards from $r=R_{\rm out}$ to $r=R_{\rm adv}$ to give
\begin{equation}
\label{anaeq}
{T_{\rm adv}-T_d\over T_d}  = \left({\gamma-1\over\gamma}\right) \left({GM_p\mu\over k_BT_d R_{\rm out}}\right) \left[{R_{\rm out}\over R_{\rm adv}}-1\right],
\end{equation}
where we take the temperature at $R_{\rm out}$ to be the disk temperature $T_d$, and $T_{\rm adv}$ is the temperature at the inner edge of the entropy advection zone $r=R_{\rm adv}$. 
In terms of the Bondi radius, $R_B=GM_p/c_s^2$, 
\begin{equation}\label{eq:deltaT_adv}
{T_{\rm adv}-T_d\over T_d}  = (\gamma-1) \left({R_B\over R_{\rm out}}\right) \left[{R_{\rm out}\over R_{\rm adv}}-1\right].
\end{equation}
The factor $R_B/R_{\rm out}$ is either unity when the Bondi radius is smaller than the Hill radius (then $R_{\rm out}=R_B$), or larger than unity when the Bondi radius is larger than the Hill radius (then $R_{\rm out}=R_H$). 

Equation (\ref{eq:deltaT_adv}) shows that entropy advection can easily result in tens of percent increase in the outer boundary temperature, and possibly substantially more depending on the depth to which the flow penetrates and whether the Bondi radius is larger than the Hill radius. How will this impact the cooling luminosity? We can estimate the effect on the luminosity if we assume that the radiative zone is isothermal {(which applies only to low metallicity envelopes)}, so that the temperature at the radiative-convective boundary (RCB) is the same as the outer boundary temperature, $T_{\rm RCB}=T_{\rm adv}$. The luminosity, set at the RCB, is
\begin{equation}
    L\approx {64\pi GM_p\sigma T_{\rm RCB}^4 \nabla_{\rm ad}\over 3\kappa P_{\rm RCB}}\propto {T_{\rm RCB}^4\over \kappa P_{\rm RCB}}
\end{equation}
\citep{arras,piso} (again we will neglect the self-gravity of the envelope). For a power law dependence of opacity $\kappa\propto \rho^aT^b\propto P^a T^{b-a}$, we find that for a fixed internal entropy so that $P_{\rm RCB}\propto T_{\rm RCB}^{\gamma/(\gamma-1)}$, the luminosity varies with RCB temperature as
\begin{equation}
    {d\ln L\over d\ln T_{\rm RCB}} = 4 - b+a - (1+a){\gamma\over \gamma-1}.
\end{equation}
Taking the values $a=2/3$ and $b=3$ for molecular opacity from \cite{belllin}, we find that a change in RCB temperature $\Delta T_{\rm RCB}$ leads to a change in luminosity
\begin{eqnarray}
    {\Delta L\over L} &\approx& -4 {\Delta T_\mathrm{RCB}\over T_\mathrm{RCB}}\approx -4{T_\mathrm{adv}-T_d\over T_d}\nonumber \\
    &=& -{8\over 5} \left({R_B\over R_{\rm out}}\right) \left[{R_{\rm out}\over R_{\rm adv}}-1\right].
\end{eqnarray}
A hotter outer boundary leads to a lower luminosity. For $R_{\rm adv}/R_{\rm out}=0.3$ ($0.1$) and assuming $R_{\rm out}=R_B$, the predicted decrease in luminosity is a factor of $\approx 4$ ($\approx 14$). For dust opacity, \cite{belllin} give $a=0$ and $b=1/2$ (metal grains) or $b=2$ (ice grains). In this case, $d\ln L/d\ln T_{\rm RCB}=(1/2) - b$, so the change in luminosity will be smaller than for molecular opacity.

Since the luminosity sets the cooling timescale and the envelope mass increases as a power law in time, we expect similar changes in the time to reach crossover mass or in the envelope mass at a given time. For example, using the scalings derived by \cite{lee2} for the envelope mass at a fixed time (their eq.~17), we find $ M_{\rm env}/M_p\approx -(3/2)\Delta T_{\rm RCB}/T_{\rm RCB}$ for $a=2/3$ and $b=3$. For dust with $a=0$, the scaling is $M_{\rm env}/M_p\approx (1/4)(1-2b)\Delta T_{\rm RCB}/T_{\rm RCB}$.
The overall effect of entropy advection therefore depends on the opacity regime at the RCB and dust content of the envelope. We present the results of our more detailed models in the next section.


\begin{figure*}
\begin{centering}
	\includegraphics[scale=0.31]{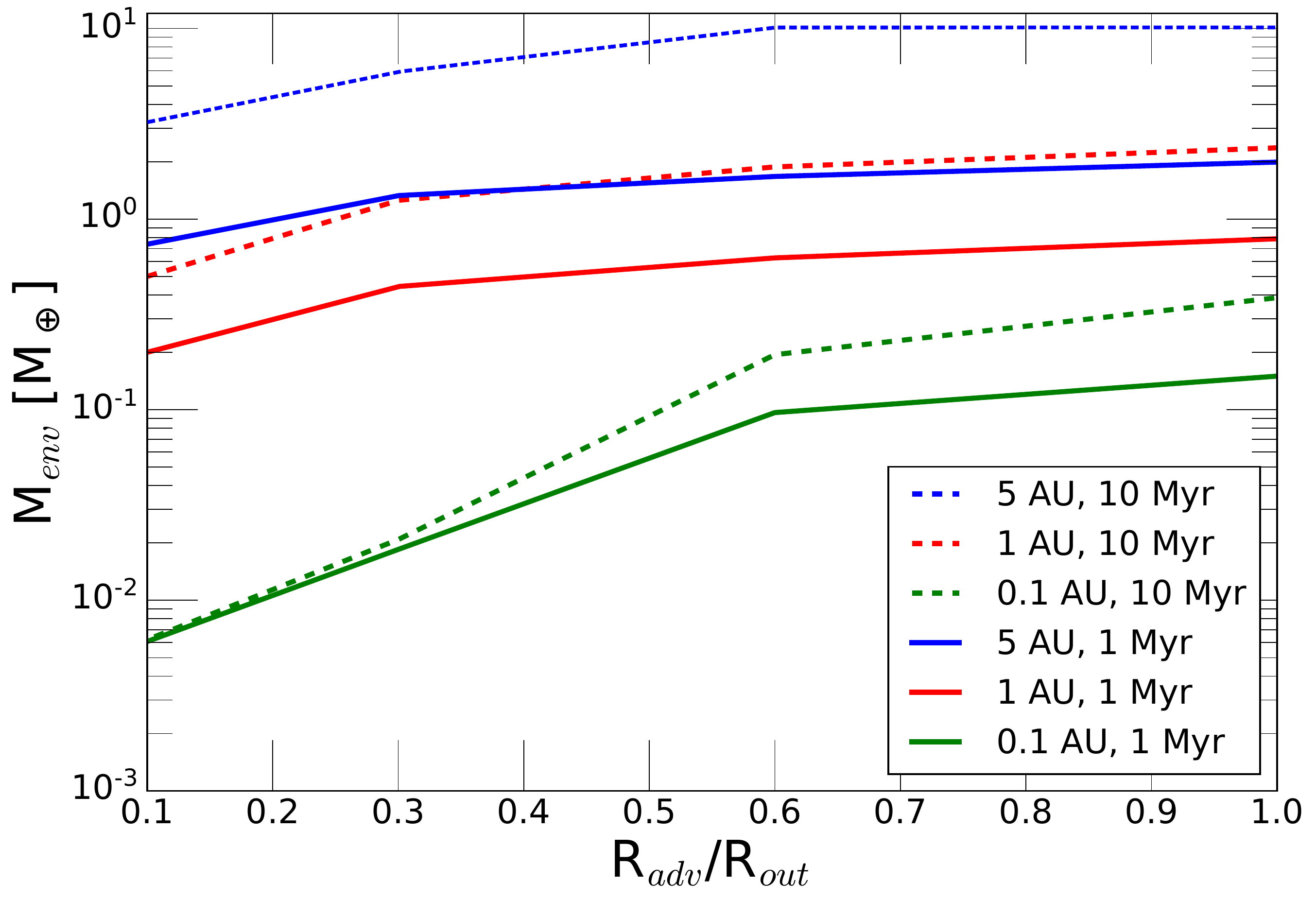}
	\includegraphics[scale=0.31]{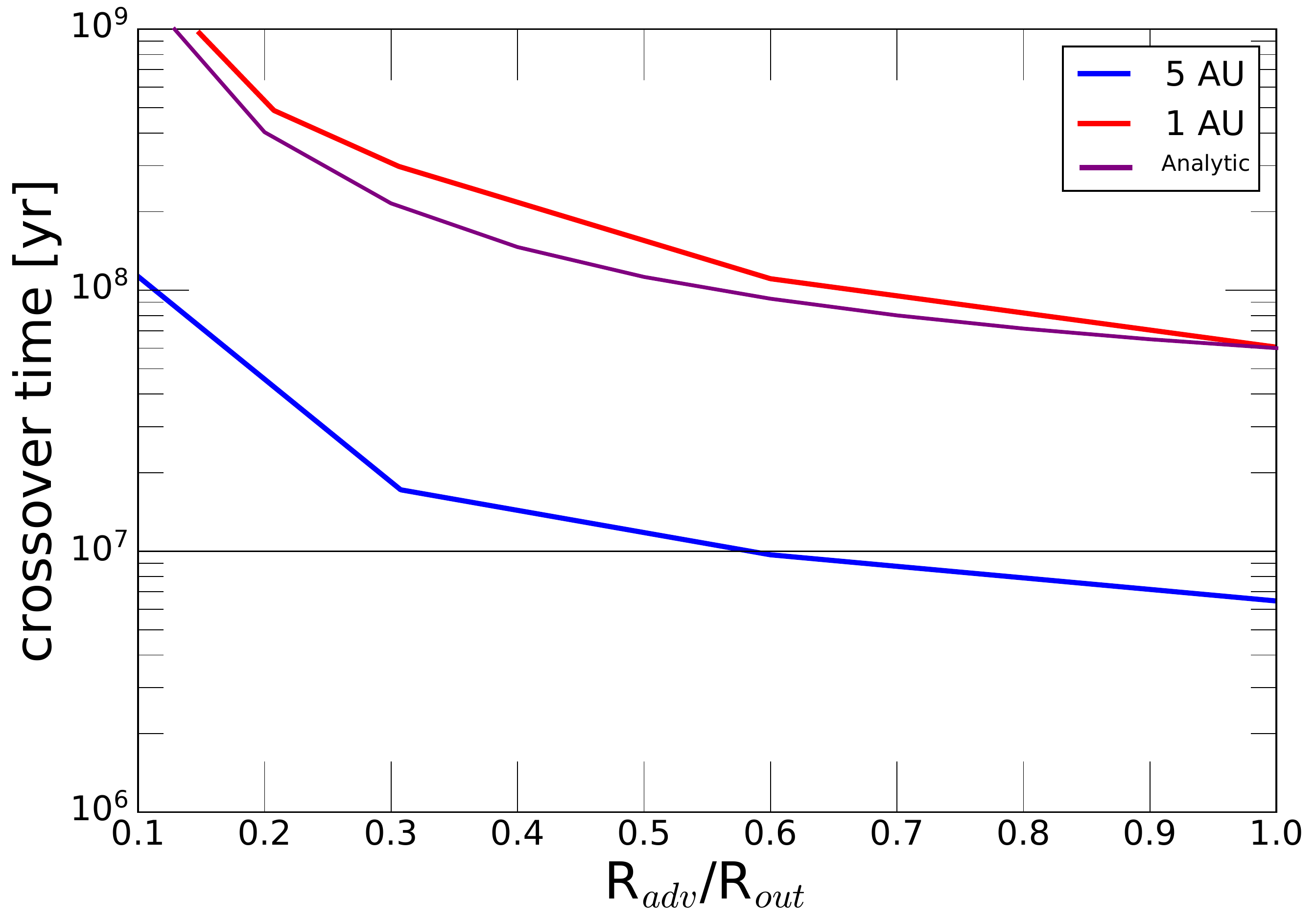}

   \caption{Left panel: The envelope mass after 1 and 10 Myr as a function of entropy advection depth $R_{\rm adv}$/$R_{\rm out}$ and the semi major axis. Right panel: The envelope's crossover time as a function of the same parameters. Here 0.1 AU is not shown since it is higher up on the plot. In both panels, we use the radiative disk model. The violet curve is obtained from our analytical equation (\ref{anaeq}), and assuming that $t_{\mathrm{co}} \propto T_d^{2.5}$, and that $T_\mathrm{RCB}\propto T_d$. }
    \label{fig:radv}
    \end{centering}
\end{figure*}

\begin{figure*}
\begin{centering}
	\includegraphics[scale=0.25]{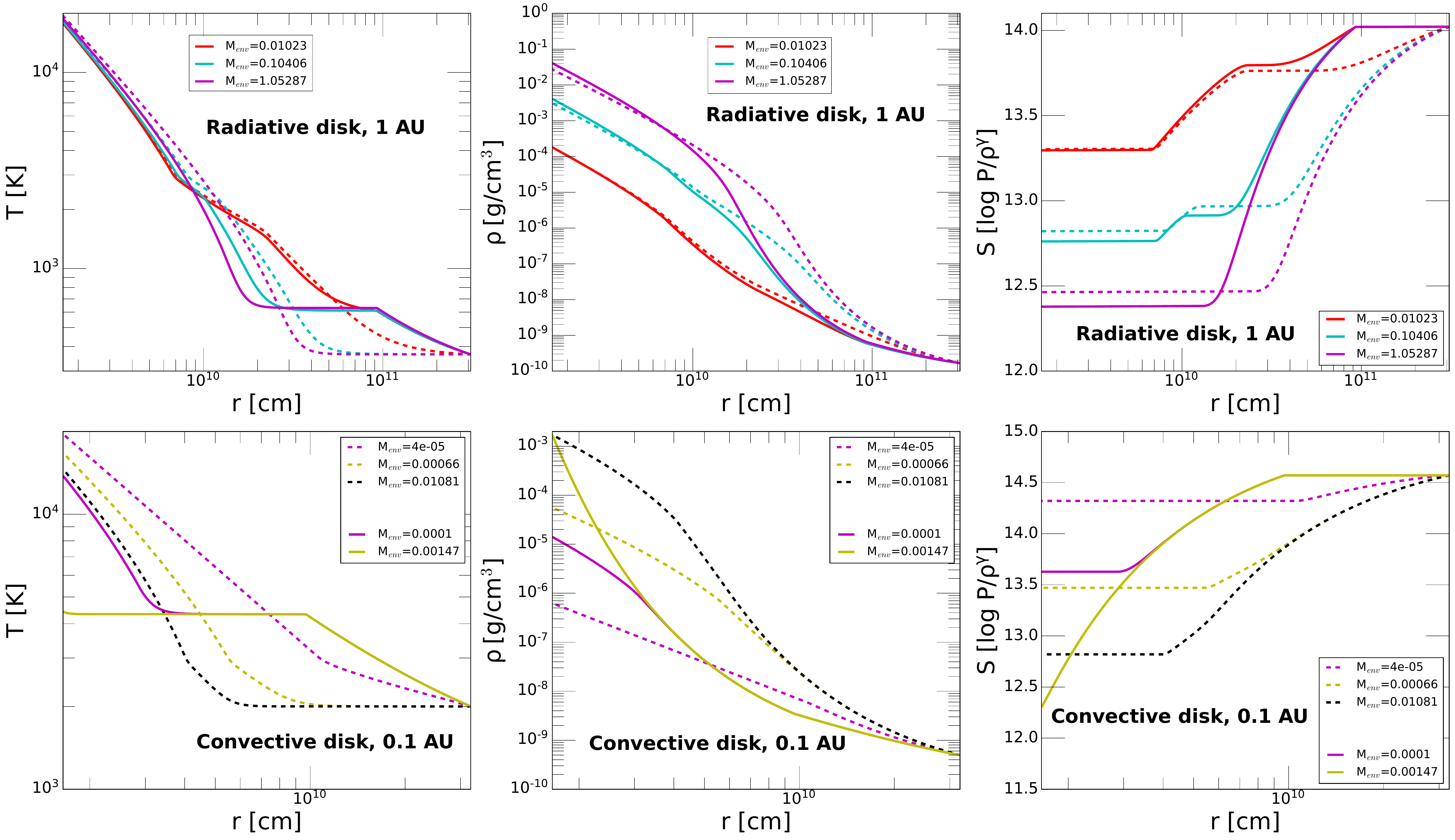}
   \caption{Upper panels: The envelope's temperature, density, and entropy radial profiles for a core at 1 AU of a radiative disk as a function of the envelope's mass. Dashed and solid lines represent respectively the cases without and with entropy advection. Lower panels: Same as above, but for a core at 0.1 AU in a convective disk. Here, for the case with advection, only two envelope masses are shown since the system quickly becomes fully isothermal and stops cooling, halting gas accretion entirely. }
    \label{fig:dt}
    \end{centering}
\end{figure*}

\begin{table}
\begin{tabular}{l|c|l|c|l|}
\cline{2-5}
                             & \multicolumn{2}{l|}{Radiative disk}                    & \multicolumn{2}{l|}{Convective disk}                                         \\ \hline
\multicolumn{1}{|l|}{}       & \multicolumn{1}{l|}{$T_d$ {[}K{]}} & $\rho_d$ {[}g/cm$^3${]} & $T_d$ {[}K{]} & $\rho_d$ {[}g/cm$^3${]}                                            \\ \hline
\multicolumn{1}{|l|}{20 AU}   & 24                             & 1.18$\times 10^{-12}$ & 21        & 1.32$\times 10^{-13}$   \\ \hline
\multicolumn{1}{|l|}{5 AU}   & 85                             & 1.17$\times 10^{-11}$ & 80        & 1.13$\times 10^{-12}$                                            \\ \hline
\multicolumn{1}{|l|}{1 AU}   & 365                            & 1.67$\times 10^{-10}$ & 382       & 1.38 $\times 10^{-11}$ \\ \hline
\multicolumn{1}{|l|}{0.1 AU} & 2000                           & 7.45$\times 10^{-9}$  & 2000      & 4.96$\times 10^{-10}$                                            \\ \hline
\end{tabular}
\caption{The disk temperature and density values used for our simulations. For the disk model parameters, we take $\dot M=3\times 10^{-8}\ M_\odot\ {\rm yr^{-1}}$, $\alpha=10^{-2}$, and $M_\star=M_\odot$.}
\label{table}
\end{table}

\subsection{Results}

We now use our numerical models to evaluate the influence of the disk conditions on the envelope growth for a fixed mass $10\ M_\oplus$ core at three different locations in the disk, at $0.1$, $1$ and $5\ {\rm AU}$. The disk density and temperature at these locations for the different models are given in Table \ref{table}. For the purposes of these calculations, we assume that the disk conditions are fixed, i.e.~the disk accretion rate is constant in time. As discussed above, dust opacity is suppressed by a factor of 10 for these models. 

\subsubsection{The effect of entropy advection on envelope growth}

Considering the control case of a radiative disk without entropy advection, we overall find the expected trends, where the envelope's luminosity decreases with time, the RCB moves to higher pressures, and the envelope mass increases. Additionally, the envelope mass increases with the semi-major axis as the lower ambient temperatures allow for more efficient cooling. 

In Fig. \ref{fig:radv}, we show the envelope mass at 1 Myr as a function of the semi major axis and the depth of penetration of the disk flow inside the envelope, parametrized through $R_{\rm adv}$/$R_{\rm out}$. All simulations were done in our nominal radiative disk. While non-isothermal hydrodynamic simulations found the flow to penetrate only to $\sim 0.3$ $R_H$, we try values down to 0.1 $R_H$ to explore the parameter space. At 5 and 1 AU, entropy advection introduces a difference in envelope mass of a factor $\sim$ 2--3 for $R_{\rm adv}$/$R_{\rm out}$=0.3. This indicates that this mechanism does not affect the cooling rate considerably for advection at the level reported by hydrodynamic simulations, even after 1 Myr. If we assume that the disk flow can penetrate to 0.1 $R_{\rm out}$ however, then the difference is  larger ($\sim$ 4). 

At 0.1 AU, the effect of entropy advection is much more dramatic. We find an order of magnitude difference in envelope mass for $R_{\rm adv}$/$R_{\rm out}$=0.3, and then an even larger effect for deeper advection. As we discuss more in section \ref{cdisks}, this is because the envelopes become fully-radiative, lengthening their cooling times considerably.


To understand why entropy advection has a mild effect for $R_{\rm adv}$/R$_{out}\geq$ 0.3 at 1 and 5 AU, we show the evolution of the radial gas temperature, density, and entropy profiles for the two cases with $R_{\rm adv}$/$R_{\rm out}$= 1 and 0.3 at 1 AU in the upper panels of Fig. \ref{fig:dt}. Curves of the same color correspond to the same envelope mass. While the disk flow is limited to the outer envelope, the density plots show that most of the envelope's mass is concentrated in the inner envelope close to the core (as also noted by \cite{lee2}). Even though the profiles shapes for the two cases differ in the outer envelope, the differences are less pronounced close to the core where the gas is concentrated.
Another way to interpret these results is by considering equation (24), and assuming that $T_\mathrm{RCB}\propto T_\mathrm{adv}$, and that the cooling rate of the envelope is controlled mainly by $T_\mathrm{RCB}$. Equation (24) shows that for $R_{\rm adv}/R_\mathrm{out}=$ 0.3, $T_\mathrm{RCB}$ is increased only mildly.  

\begin{figure}
\begin{centering}
	\includegraphics[scale=0.35]{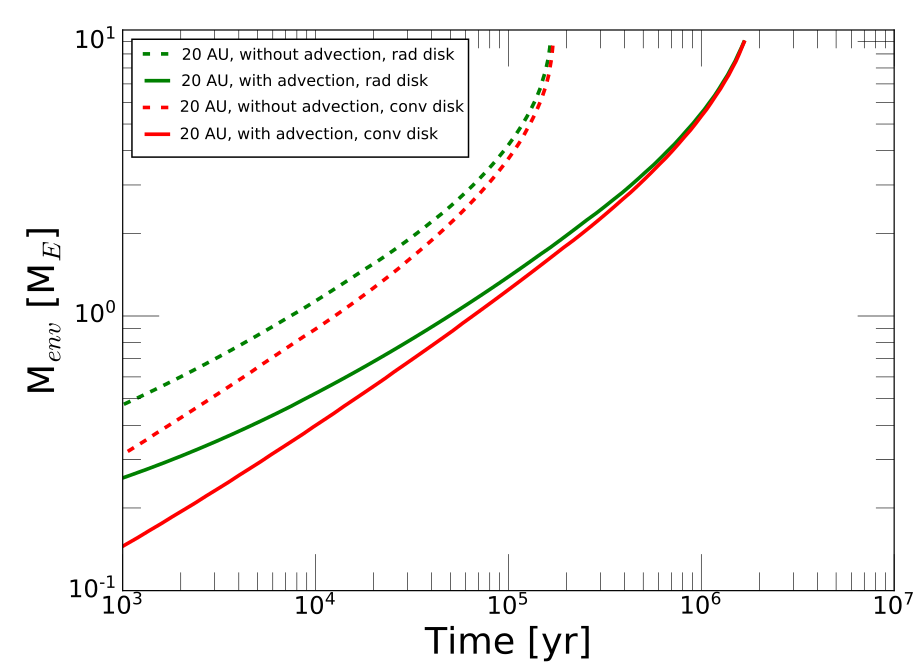}
   \caption{Envelope mass as a function of time for a 10 M$_\oplus$ core at 20 AU, as appropriate for the formation of Uranus. We show results for both radiative (green curves) and convective disks (red curves), with (solid) and without (dotted) entropy advection. In both disks, entropy advection delays runaway accretion by one order of magnitude in time. }
    \label{fig:uranus}
    \end{centering}
\end{figure}

It is important to emphasize however than this small factor of $\sim 2$--$3$ envelope mass difference for any given time introduced by entropy advection at large disk radii can delay runaway accretion. 
To illustrate the potential consequences on the time to reach runaway gas accretion, consider a case where the lifetime of the gas disk is $10\ {\rm Myr}$, indicated by the horizontal line in the right panel of Fig.~\ref{fig:radv} which shows the time to reach crossover $t_\mathrm{co}$ (at which the envelope mass equals the core mass). Considering the case at 5 AU after 10 Myr, the envelope reaches crossover mass and runaway accretion for any $R_{\rm adv}$/$R_{\rm out}$ larger than $\sim$ 0.6. Entropy advection for $R_{\rm adv}$/$R_{\rm out}$=0.3 is hence able to delay runaway accretion beyond the dissipation of the disk.  We find an even greater effect further out where the disk is colder. Figure \ref{fig:uranus} shows models at 20 AU, as appropriate for formation of Uranus. {The convective and radiative disks show very similar behavior; in each case, including entropy advection with $R_{\rm adv}/R_{\rm out}=0.3$ delays runaway by an order of magnitude (from $\approx 2\times 10^5$ yrs to $\approx 2$ Myrs for the parameters chosen here), even though in no case runaway time is delayed beyond the lifetime of the disk (due to its low entropy at 20 AU).}


\begin{figure*}
\begin{centering}
	\includegraphics[scale=0.30]{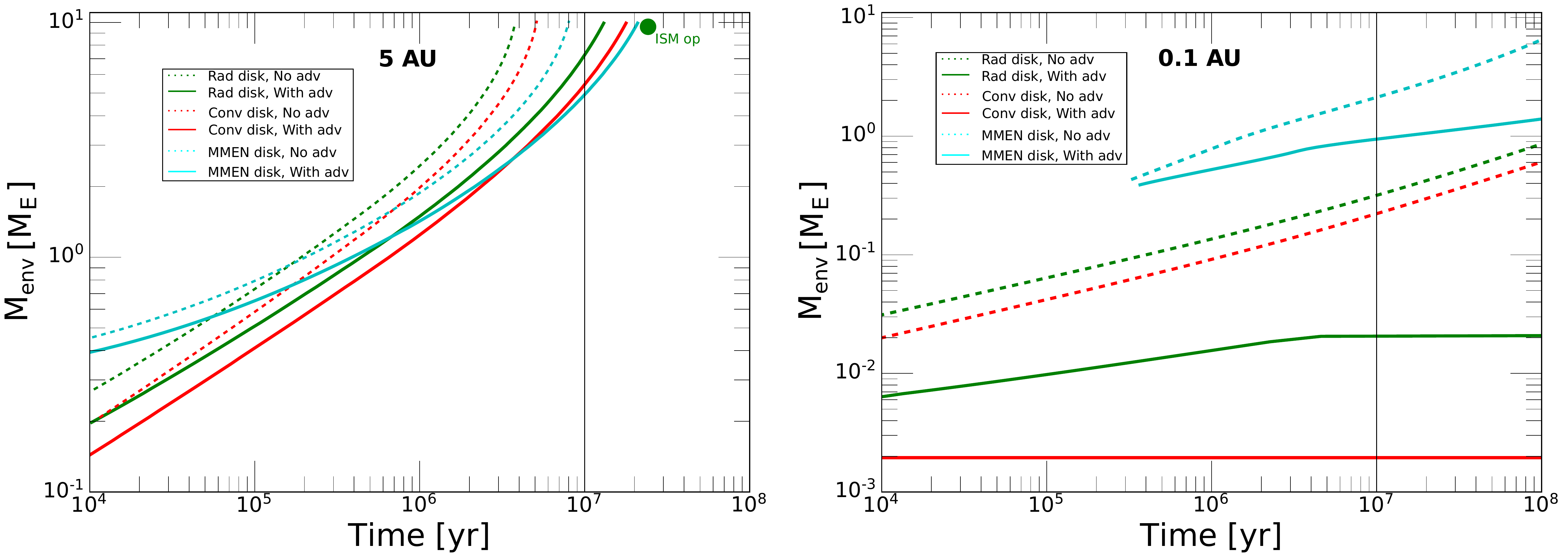}
   \caption{Envelope mass for a $10\ M_\oplus$ core as a function of time at 5 AU (left panel) and 0.1 AU (right panel). We show results for radiative (green), convective (red), and MMEN (cyan) disks, and with (solid) and without (dotted) entropy advection. For radiative and convective disks at 0.1 AU (red and green solid curves), the high disk entropy forces the envelope into a fully isothermal (thus radiative) state, shutting down its cooling and mass accretion. The green dot in the left panel shows effect on runaway time (radiative disk, without entropy advection) of using the full ISM dust opacity (other cases have dust opacity reduced by a factor of 10, see text for discussion).}
    \label{fig:radvsconv}
    \end{centering}
\end{figure*}

\subsubsection{The effect of disk entropy on envelope growth}
\label{cdisks}

In Fig. \ref{fig:radvsconv} we investigate the effects of the disk entropy by comparing the envelope cooling time between radiative and convective disks, at 0.1 and 5 AU. We moreover show the results obtained using the MMEN for comparison.

Considering the cases at 5 AU, we find again a relatively modest effect. Comparing independently the cases with and without entropy advection (solid and dashed lines respectively) at a time 1 Myr, we see that the envelope mass in the convective disk is approximately a factor of 1.5 lower than in radiative disks. This is due to the higher ambient (disk) entropy in convective case as can be seen in Fig. \ref{fig:disk1}, implying simultaneously roughly the same disk temperature, but a significantly lower disk density (Fig. \ref{fig:disk1}). As previously, we see that entropy advection acts to delay runaway. In all three disk models (radiative, convective, and MMEN), runaway accretion is reached within our fiducial disk lifetime of 10 Myr without entropy advection, but including entropy advection with $R_{\rm adv}$/$R_{\rm out}$ = 0.3 however precludes runaway within $10\ {\rm Myr}$.

At $0.1\ {\rm AU}$, however, the disk model has a much more dramatic effect when entropy advection is included. First considering the case at 0.1 AU without entropy advection, the envelope mass in the convective disk is again a factor 1.5 lower than for the radiative disk for both 1 and 10 Myr. 
The influence of the disk model is much greater when entropy advection is included. 
The solid green and red line in Fig.~\ref{fig:radvsconv} represents the case of radiative and convective disks with entropy advection, at 0.1 AU. The behavior of the atmosphere here is clearly different than the rest. The envelope mass plateaus quickly and the system evolution slows down significantly to a steady state. The final envelope mass is orders of magnitude less massive than the corresponding case without entropy advection.

The stalling of the cooling of the envelope is due to the fact that it becomes fully-radiative. This can be seen in Fig. \ref{fig:dt} (bottom panels), where we plot the evolution of the entropy profile of the envelope for the convective disk case. The entropy of the envelope is always lower than that of the disk, as physically our setup is equivalent to integrating inward from the disk entropy to an internal adiabat, passing through a radiative zone where entropy decreases. For low envelope mass (earlier times, higher luminosity), the envelope is characterized by an outer adiabatic plateau due to entropy advection, then a radiative zone where the entropy decreases, then finally the inner adiabat in the convective zone where the entropy is constant again. This inner convective zone then becomes smaller with increasing envelope mass (increasing time and decreasing luminosity), till it disappears entirely creating a fully radiative envelope inside the outer adiabatic zone. Once this stage is reached, the envelope cooling slows down dramatically, as the entropy profile stops evolving, and orders of magnitude decrease in the luminosity are needed to jump into the next envelope snapshot with slightly higher mass. 

To understand this further, we also plot the density and temperature evolution for this case in Fig. \ref{fig:dt}. Due to the high temperature but low density, the radiative gradient $\nabla_\mathrm{rad}\propto T^{-4}$ is vanishingly small, leading to an almost isothermal radiative zone in the envelope, with the temperature increasing adiabatically in the outer adiabatic zone, and (for low envelope masses) again in the inner convective region. Once the inner convective zone has disappeared however, the entire envelope inside the outer adiabat becomes isothermal, and thus no more cooling is possible since the temperature gradient has vanished. This also freezes the density profile, and therefore the total mass. This same type of behavior for isothermal envelopes was also noted by \cite{lee2}. 


Looking at the cases of the MMEN disk at 0.1 AU, we find that the envelope grows significantly more than in the radiative and convective disks. This is true both with and without entropy advection. The larger envelope mass is due to the significantly lower entropy of the MMEN compared to ours as seen in Fig. \ref{fig:disk1}, allowing for a more efficient cooling. Physically, this is because MMEN disks are passive, while our radiative disk is viscously heated. Note that in the MMEN disk at 0.1 AU the envelope does not reach runaway accretion in 10 Myr. This is in contrast with the results of \cite{lee1}, but we find that this is due to the differences in the model details, including the opacity prescription and equation of state. {\cite{lee1} use a generally lower adiabatic $\gamma$ than our 1.4, that can go down to $\sim 1.2$ due to their EoS accounting for the effects of H$_2$ dissociation. We are able to reproduce their results by adopting a lower value of $\gamma$ in our model.} 

\begin{figure}
\begin{centering}
	\includegraphics[scale=0.32]{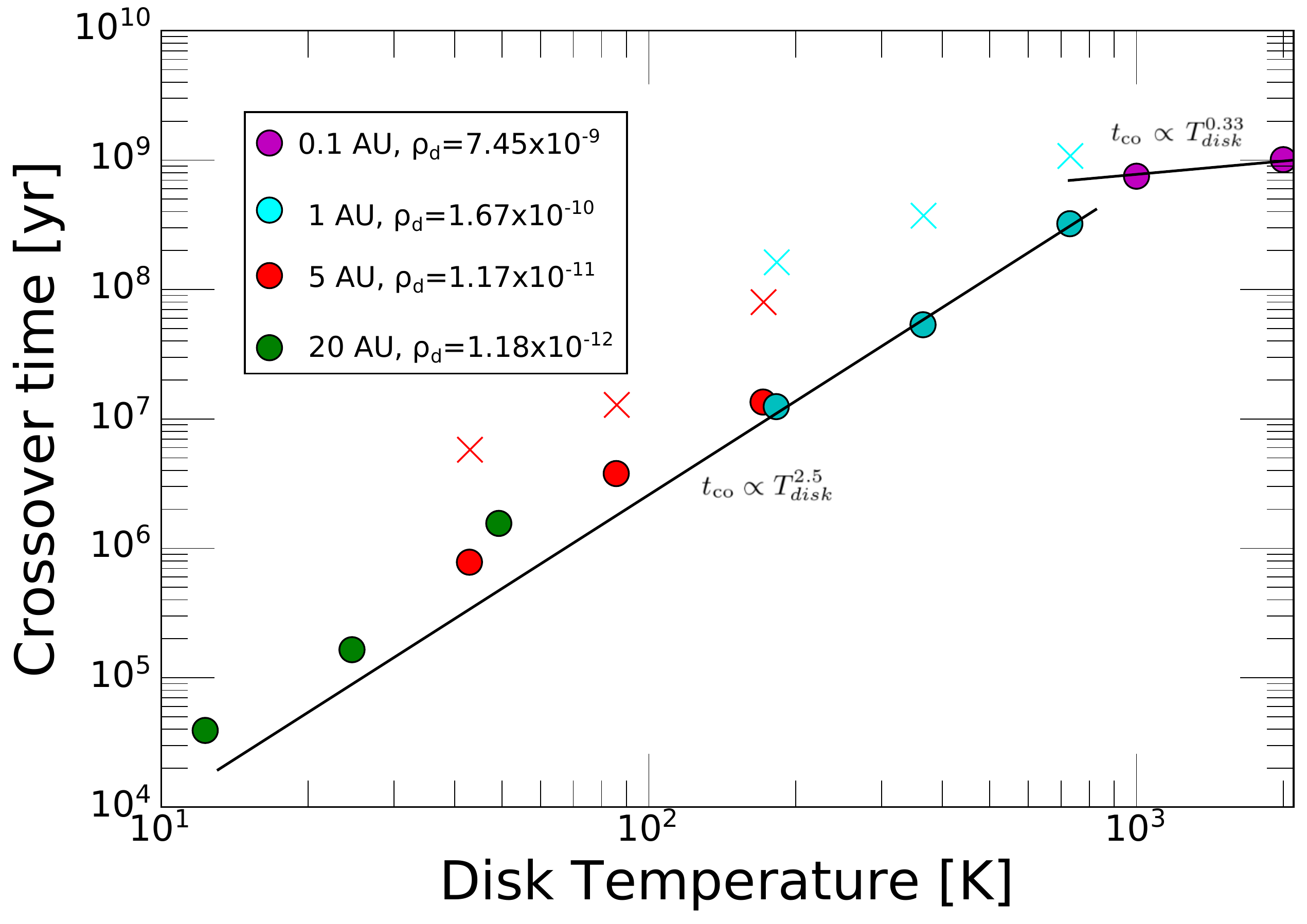}
   \caption{The envelope crossover time for a fixed core mass of $10 M_\oplus$ as a function of the radiative disk's temperature and density. Circles and X represent respectively models without and with entropy advection. Circles of the same color have the same semi major axis (and thus density), but differ in temperature by a factor 2. The steep dependency slope of the crossover time on the disk temperature shallows out significantly at 0.1 AU. }
    \label{fig:dentencp}
    \end{centering}
\end{figure}

\cite{lee1} earlier studied the dependence of crossover time on the disk temperature and density, and found that the disk conditions do not strongly affect the time to crossover. 
To investigate this in our models, we take the radiative disk case, and, at each radius, run two extra simulations with the temperature artificially scaled by factors of 0.5 and 2. Our results are shown in Fig. \ref{fig:dentencp}. At 0.1 AU we retrieve the same $t_{\mathrm{co}} \propto T_d^{0.33}$ dependency as \cite{lee1}. At lower temperatures, however, this relation gets much steeper, between $t_{\mathrm{co}} \propto T_d^2$ and $t_{\mathrm{co}} \propto T_d^{2.5}$, similar to what we get from the analytical treatment of \cite{piso} (using their eqs.~[1b] and [34]). This change in dependency is due to opacity regime change. At 100 K for example, a factor 2 in temperature can change the opacity regime entirely from grains-dominated to molecular gas-dominated due to ice sublimation. At 1000 K however, as long as hydrogen has not dissociated, a factor of two change in temperature will keep the system in the same opacity regime. When considering the density, we retrieve the same $t_{\mathrm{co}} \propto \rho_d^{-0.2}$ power law as \cite{lee1}, since our opacity laws depends significantly more on temperature than on density.

\section{Summary \& discussion}

In this paper we considered the cooling and subsequent growth of the envelope of a $10 M_\oplus$ core embedded in a protoplanetary disk. We aimed to understand how the long term envelope cooling is affected by two different aspects of the disk: 1) the disk entropy and thermal state, and 2) advection of entropy into the atmosphere by disk flows (``entropy advection''). We considered three different disk models: accreting disks with vertical energy transport by radiation or by convection, and a disk heated by irradiation from the central star (the minimum mass exoplanetary nebula MMEN), and modelled the envelope growth at different locations in the disk. 

Our main findings are as follows:
\begin{itemize}
\item For cores located at $1$ and $5\ {\rm AU}$, entropy advection to the depth $R_{\rm adv}$/$R_{\rm out}=0.3$ (as found in non-isothermal hydrodynamic simulations) decreases the envelope mass at any given time by only a factor of $\sim 2$--$3$. This is true for both radiative and convective disks. These cores are able to develop massive envelopes of the order of $\sim M_\oplus$ even when entropy advection is included. The runaway time is delayed by a factor $\sim 3$--$5$, possibly pushing it beyond the disk lifetime.
\item For both radiative and convective disks at 0.1 AU, we find that the disk entropy is high enough for entropy advection to force the envelope into a fully-radiative state after a brief period of cooling. Once this has been reached, the envelope growth stalls entirely as it becomes isothermal, keeping the envelope mass orders of magnitude lower than the corresponding case without entropy advection.  
\item The reduction in envelope mass by entropy advection in the inner parts of the disk is significantly more pronounced in convective disks due to their higher entropy, leading to steady-state envelope mass of only $\sim 2\times 10^{-3}$ M$_\oplus$, compared to $\sim 2\times 10^{-2}$ M$_\oplus$ in radiative disks (Fig.~\ref{fig:radvsconv}).
\item When irradiation is the only source of energy heating the disk, as assumed in many previous models (e.g.\citealt{chiang,piso,rafikov2006}), the temperature and entropy close to the star ($\sim 0.1\ {\rm AU}$) are significantly lower than in accreting disks. This leads to much faster envelope growth. There is a large difference in the inner parts of the disk despite similar conditions further out because irradiated disks have an entropy profile that increases outwards, as compared to the relatively shallow entropy gradient in viscously heated disks (Fig.~\ref{fig:disk1}).
\end{itemize}

{Current Super Earths population is consistent with a wide range of chemical compositions for both the cores and envelopes, and, by consequence, envelope mass fractions \citep{lopez}. The low envelope to core mass ratios we obtain at 0.1 AU with entropy advection ($\sim 10^{-4}$--$10^{-2}$) are consistent with a volatiles-poor subpopulation, while our no-advection cases are consistent with relatively volatiles-rich Super Earths.}
Moreover, the less massive envelopes we obtain are especially prone to photoevaporation \citep{ow1,ow2}. Our results hence point towards some of the high mass, small radius Kepler planets being naked cores, and provide a natural formation mechanism for such planets. These cores however could regenerate a tenuous but detectable atmosphere via geochemical processes, as in 55 Cancri e \citep{55cnce}. {Note however that this does not take into account the envelope's mass loss due to its own cooling luminosity after the disk's dispersal \citep{ginz,gupta}.}

The slowing of envelope growth by entropy advection could be important for core accretion models of Uranus and Neptune, which face the problem of needing to grow the solid core quickly enough, while at the same time avoiding runaway gas accretion and growth to a gas giant \citep{helled2014}. Our calculations do not take into account growth of the solid core, and so we cannot construct a detailed model of the formation of Uranus or Neptune. We did, however, calculate the growth of the envelope at an orbital radius of 20 AU, appropriate for Uranus, to illustrate the effect of entropy advection (Fig.~4). The delay in runaway time may help with models of ice giant formation. Note that our models do not include the timescale to build the core, which could be the rate-limiting step if the growth is by planetesimal accretion, or could be quite rapid in the case of pebble accretion or if the cores form closer in and are scattered outwards \citep{helled2014}. Slowing the envelope growth means that the planet spends more time with envelope masses of $\sim M_\oplus$ needed to explain the inferred H/He fractions in Uranus or Neptune. This may alleviate the fine-tuning of the disk lifetime needed to reproduce the gas fractions of the ice giants (e.g.~see discussion in \citealt{helled2019}). 

Our results suggest that energy transport by convection in the protoplanetary disk could {be partially responsible for the low envelope masses of super-Earths.} Whether or not protoplanetary disks become convective is an open question. \cite{lin1} derived a simple criteria for triggering vertical convective instability in disks: for a power law opacity $\kappa \propto T^\beta$: $\nabla_{\rm ad} \geqslant 1/(4-\beta)$. For $\nabla_{\rm ad}\sim 0.285$, and the opacity prescription of \cite{belllin}, this condition can be satisfied in large parts of the disk, including at 0.1 AU. While these models were set aside for a while due to conflicting numerical results \citep{klahr2007}, more recent studies, for example \cite{rafikov,lesur,fromang,klahr}, showed that vertical convective instability in disks cannot yet be excluded as a plausible heat and angular momentum transport mechanism.  

While turbulence likely plays a role in planetesimals formation \citep{johansen}, and convective disks offer that naturally through hydrodynamic turbulence, ALMA observations suggest that at least some disks might be close to laminar \citep{turbobs1,turbobs2}. These measurements, while limited to HD 163296, put into doubt the presence of a significant source of turbulence (MRI or purely hydrodynamic) in disks. {If indeed future observations rule out the presence of turbulent (and specifically convective) disks, then models should focus on the lower entropy radiative disks.}      

Our results point to the important role of the disk in the accretion rate of protoplanetary envelopes, and the uncertainty introduced by our lack of knowledge of conditions in the disk.
For the calculations in this paper we made a number of approximations that should be relaxed in future work. We did not follow the time-evolution of the protoplanetary disk, but instead assumed constant disk conditions. We also did not follow the growth of the protoplanetary core, but instead assumed a constant core mass. Another important factor is the dust opacity, which we took to be reduced from the interstellar value by a fixed factor of 10.  Our prescription for entropy advection is simplified in particular by assuming spherical symmetry whereas the disk flows into the Hill sphere in simulations are definitely not spherically-symmetric. Further work that relaxes these assumptions will allow for a more accurate assessment of how the protoplanetary disk influences the final envelope mass of super-Earths and ice giants.






\subsubsection*{Acknowledgements}

We thank Chris Ormel, Ravit Helled, and the members of the ISSI International Team on Ice Giants  for useful discussions. We thank the referee, Eve Lee, for her useful comments that significantly improved the manuscript. The computations were performed on the Sunnyvale cluster at the Canadian Institute for Theoretical Astrophysics (CITA). M.A.-D. is supported through a Trottier postdoctoral fellowship. A.~C. is supported by an NSERC Discovery grant and is a member of the Centre de Recherche en Astrophysique du Qu\'ebec (CRAQ).








\appendix


\bsp	
\label{lastpage}
\end{document}